\documentclass[fleqn,usenatbib]{mnras}

\usepackage[T1]{fontenc}
\usepackage{ae,aecompl}

\usepackage{graphicx}   
\usepackage{amsmath}	
\usepackage{amssymb}	

\usepackage{subfigure}

\def\xmmto{{\it XMM-Newton}}
\def\xmm{{\it XMM-Newton }}
\def\cha{{\it Chandra }}
\def\suz{{\it Suzaku }}

\def\asca{{\it ASCA }}
\def\ascato{{\it ASCA}}
\newcommand{\Ka}{\ensuremath{\hbox{K}\alpha~}}
\newcommand{\Kb}{\ensuremath{\hbox{K}\beta~}}
\newcommand{\KA}{\ensuremath{\hbox{K}\alpha}}
\newcommand{\kms}{\ensuremath{\hbox{km}~\hbox{s}^{-1}}}

\title[circumstellar material of GX 301-2]{The spatial distribution of circumstellar material of the wind-fed system GX 301-2}

\author[X. Y. Zheng et al.]{
Xueying Zheng,$^{1,2}$\thanks{E-mail: zhengxy@nao.cas.cn}
Jiren Liu,$^{1}$\thanks{E-mail: jirenliu@nao.cas.cn}
Lijun Gou$^{1,2}$
\\
$^{1}$Key Laboratory for Computational Astrophysics, National Astronomical Observatories, Chinese Academy of Sciences,\\  ~Datun Road A20, Beijing 100012, China\\
$^{2}$School of Astronomy and Space Sciences, University of Chinese Academy of Sciences, Datun Road A20, Beijing 100049, China\\
\\
}

\date{Accepted XXX. Received YYY; in original form ZZZ}
\pubyear{2019}

\begin{document}
\label{firstpage}
\pagerange{\pageref{firstpage}--\pageref{lastpage}}
\maketitle

\begin{abstract}
The distribution of the circumstellar material in systems of supergiant X-ray binaries (SgXBs)
is complex and not well probed observationally. We report a detailed study of the spatial distribution of the 
Fe \KA-emitting material in the wind-fed system GX 301-2, by measuring the time delay between the 
Fe \Ka line and the hard X-ray continuum ($7.8-12$ keV) using the cross-correlation method, based on
\xmm observation. 
We found that to obtain the true time delay, it is crucial to subtract the 
underlying continuum of the Fe \Ka line.
The measured size of the Fe \KA-emitting region over the whole observation period 
is $40\pm20$~light-seconds. It is 5 times larger than the accretion radius estimated from  
a quasi-isotropic stellar wind, but consistent with the one estimated from a tidal stream, which could be the dominant mass-loss mechanism of GX 301-2 as inferred from the 
orbital distribution of the absorption column density previously.
The measured time delay of the quiescent period is 
a little smaller than those of the flare periods, revealing the unsteady behaviour of the 
accretion flow in GX 301-2. 
Statistical and detailed temporal studies of the circumstellar 
material in SgXBs are expected for a large sample of SgXBs with future
X-ray missions, such as Athena and eXTP.

\end{abstract}




\section{Introduction}

\begin{figure*} 
  \includegraphics[width=0.4\linewidth, trim=0cm 0.0cm 0cm 0cm,clip,angle=270]{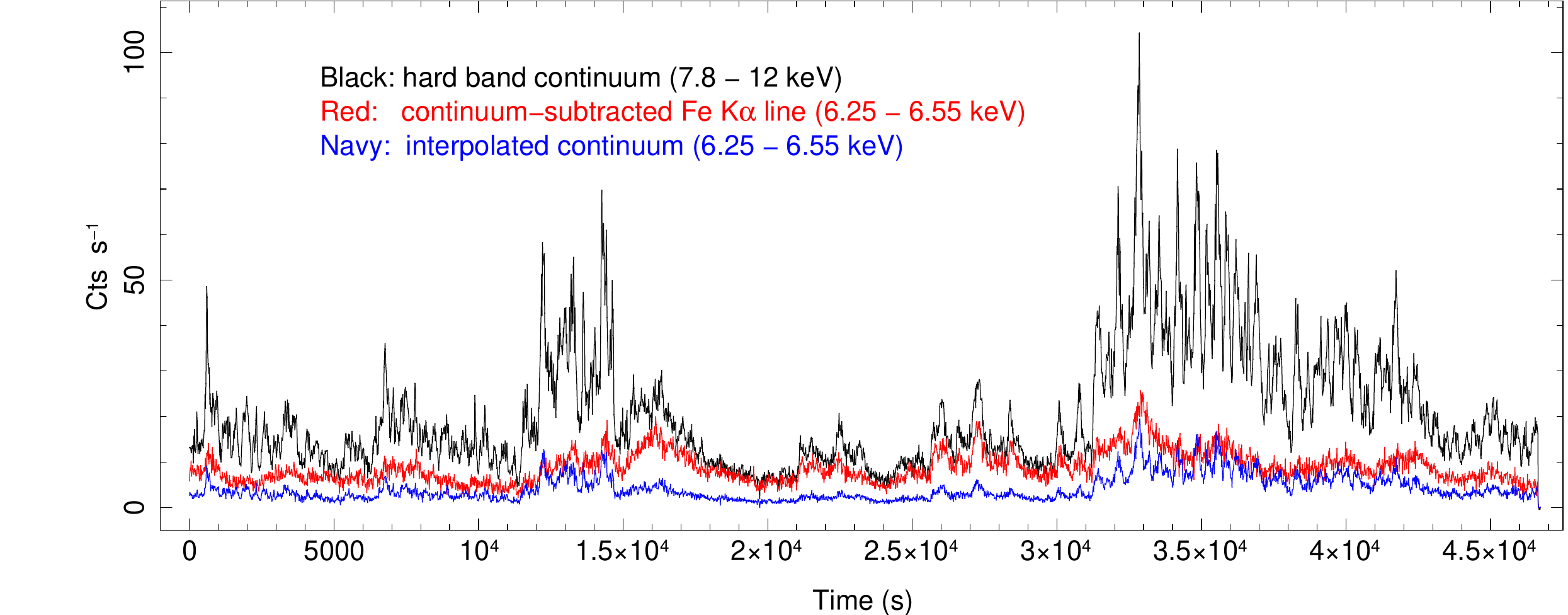}
  \caption{ \xmmto/EPIC-pn light curves of GX 301-2 for the hard continuum ($7.8-12$ keV, black), the net Fe \Ka
	  line ($6.25-6.55$ keV, red) and the interpolated continuum in the Fe \Ka band ($6.25-6.55$ keV, navy). They are binned in 10 s.
}
  \label{fig:Fe_continuum}
\end{figure*}

In systems of supergiant X-ray binaries (SgXBs), the compact star (either neutron star or black hole)
accretes the wind material from its massive companion star and emits strong X-ray radiation
\citep{Martinez-Nunez2017SSRv}. 
The compact star produces a bow shock in the surrounding material and is 
trailed by an accretion wake \citep[e.g.][]{Davidson1973ApJ}. 
The accretion flow in such a binary system is generally unsteady as shown in numerical 
simulations \citep[e.g.][]{Taam1988ApJ,Blondin1990ApJ}.
The circumstellar 
material can be probed by the soft X-ray absorption of the intrinsic X-ray emission 
\citep[e.g.][]{Nagase1989PASJ}. 
The orbital distribution of the absorption column density of some sources, such as 4U 1700-37, requires an additional tidal stream \citep{Haberl1989ApJ}.
In the case of GX 301-2, a classical SgXB, the modeled mass-loss rate of the stream exceeds that in the stellar 
wind, thus the tidal stream can dominate the evolution of the massive companion star \citep{Leahy2008MNRAS}.
The spatial distribution of the circumstellar material can be probed 
with X-ray fluorescence lines, which are produced when the surrounding material reprocesses 
the intrinsic hard X-ray emission of the compact star. 
The most prominent fluorescence feature is the Fe \Ka line at 6.4~keV, which is a hallmark of SgXBs
(\citealt{Torrejon2010ApJ}; \citealt{Gimenez2015AA}). 

There are several ways to constrain the material distribution surrounding the compact star with 
the observed Fe \Ka line.
The line width of the Fe \Ka line can be used to constrain the size of the Fe \KA-emitting region.
The measured line widths using \cha high-energy transmission grating spectrometer (HETG)
are within 400 and 4000~\kms, indicating sizes from $10^9$ to $10^{11}$~cm, if assuming the 
Fe \KA-emitting gas is virialize \citep{Tzanavaris2018ApJ}.
For eclipsing SgXBs, the decrease of the Fe \Ka fluxes during the eclipse can be used to 
infer the Fe \KA-emitting region. 
It was found that the bulk of the Fe \Ka line originates from a region smaller than the radius 
of the donor star \citep[see][and references therein]{Martinez-Nunez2017SSRv}.

While for non-eclipsing SgXBs, it is generally hard to estimate their Fe \KA-emitting region accurately. 
This is especially true for GX 301-2, the fluorescence region of which is of
vigorous debate, although it shows one of the brightest Fe \Ka line among SgXBs.
Using \asca data of GX 301-2, \citet{Endo2002ApJ} measured a line width of $40-80$~eV for the Fe \Ka line
and inferred an emission region within $\sim10^{10}$~cm (0.3~lt-s) from the neutron star. 
\citet{Furst2011AA} investigated the cross-correlation between the continuum
($7.3-8.5$~keV) flux and the Fe \Ka line ($6.3-6.5$~keV) with \xmm data, 
and could not find a significant time delay above 2~s, implying a distance smaller than 2~lt-s. 
They also proposed a second absorber on 2000~lt-s scale from an interval of low flux 
(lasted $\sim4.5$~ks),
during which the Fe \Ka line was clearly visible, while the continuum decreased significantly.
By contrast, \citet{Suchy2012ApJ} inferred a distance greater than 700~lt-s,
based on the flat pulse profile of the Fe \Ka line using \suz observations of GX 301-2.
They attributed the lack of pulsation in the Fe \Ka line to the smearing effect of a large distance.

The absorption column density of GX 301-2 near periastron is quite high, 
$\sim2\times10^{24}~$cm$^{-2}$ \citep[e.g.][]{Furst2011AA}. Such an high column density was also 
inferred from the Compton shoulder of the Fe \Ka line by \citet{Watanabe2003ApJ}.
The different estimations of the Fe \KA-emitting region of GX 301-2 have a great impact on
the gas distribution around the neutron star.
For example, if the Fe \Ka-emitting region has a size scale $\sim1$~lt-s, a density 
$\sim7\times10^{13}$~cm$^{-3}$ is needed to provide the observed column density. 
This density is 1000 times higher than the local wind density, $6\times10^{10}$~cm$^{-3}$,
estimated from a mass-loss rate of $\rm 10^{-5}~
M_\odot$~yr$^{-1}$ and a wind velocity of 150 km\,s$^{-1}$ 
at the location of the neutron star near periastron \citep{Kaper2006AA}. 
On the other hand, if the Fe \KA-emitting region is larger than 700~lt-s, it implies that 
the Fe \Ka photons are mainly produced by some gas far away from the neutron star
and the optical companion star (the periastron distance of GX 301-2 is only 200~lt-s).
The densities of the stellar wind of GX 301-2 on such large scales is far below the observed
high column density.
Furthermore, the temporal correlation between the Fe K$\alpha$ line and the continuum,  
is apparently much shorter than 700~s \citep[e.g.][see also Figure~\ref{fig:Fe_continuum} above]{Furst2011AA}.

We note that in the cross-correlation analysis done by \citet{Furst2011AA}, they used a time 
bin of 1~s, within which the average counts of the Fe \Ka photons are $\sim$10,
and the Poisson noises could be significant. 
More importantly, they did not subtract the underlying continuum within the Fe \Ka band 
($6.3-6.5$~keV), which, as illustrated in Figure~\ref{fig:Fe_continuum}, is significant. 
As shown in our analysis, whether subtracting the underlying continuum or not totally changes the 
behavior of the cross-correlation function.
This motivates us to revisit the problem of the Fe \KA-emitting region of GX 301-2 with 
the cross-correlation method, which was not regularly exploited when estimating the spatial
distribution of the circumstellar material in SgXBs and could be potentially applied
to a large number of SgXBs with future X-ray missions of large collecting area, such as Athena and eXTP.

In \S~\ref{sec:data} we describe the data used. 
\S~\ref{sec:result} presents the analysis and results.
We discuss the implications of the results in \S~\ref{sec:discussion}.
All the errors in this paper are quoted for the confidence level of 90\%.

\begin{figure} 
  \includegraphics[width=0.7\linewidth, trim=0cm 0.3cm 0cm 0.cm,clip,angle=270]{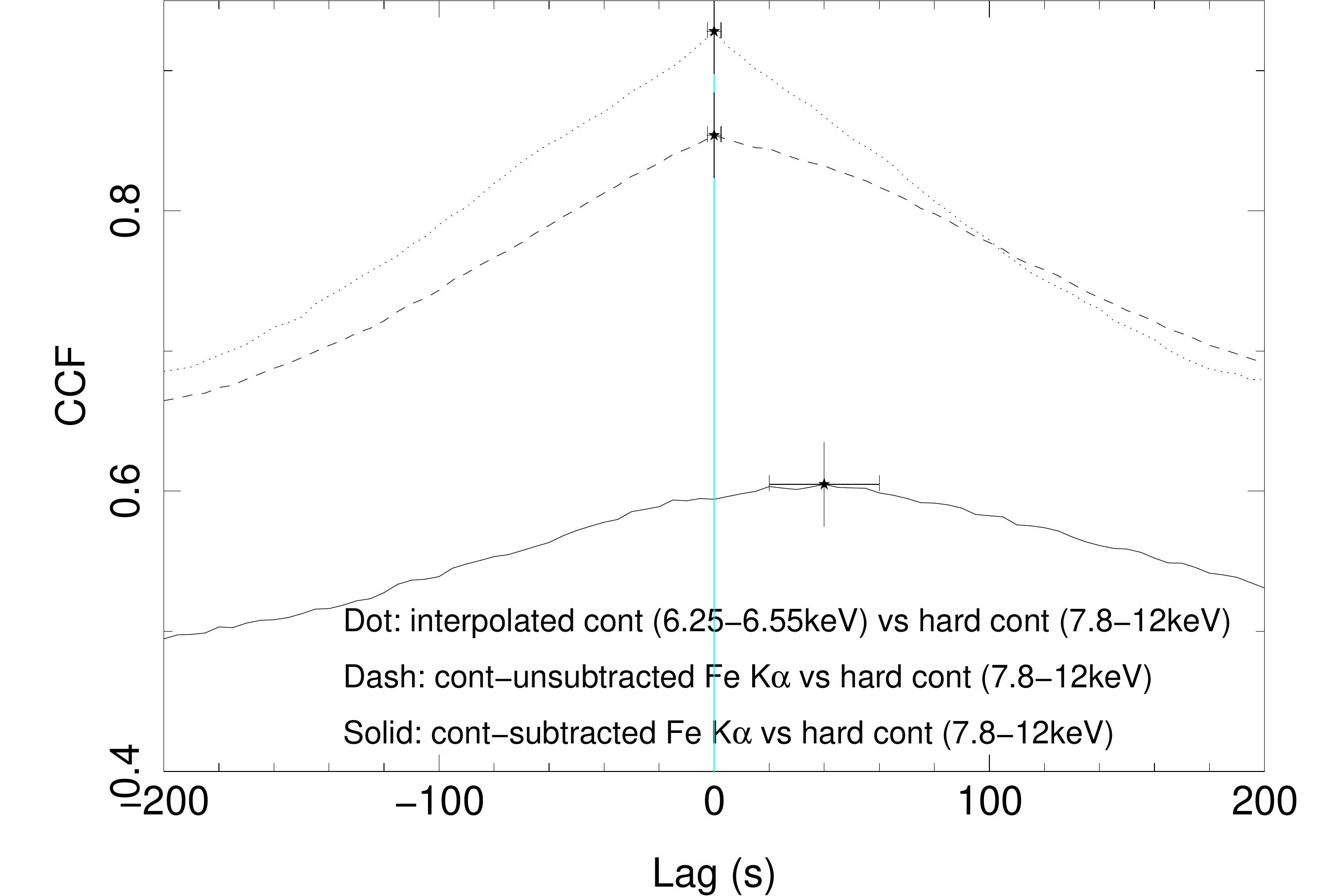} 
  \caption{CCFs between the interpolated continuum, the unsubtracted Fe \Ka line, the net Fe \Ka line,
	  and the hard continuum, respectively.
  The peaks of CCFs are showed with the vertical lines , and the error bars are estimated from re-sampling of the light curves. 
  Subtraction of the underlying continuum is crucial to reveal the true time delay of the Fe \Ka line.
  }
  \label{fig:10sbin3line}
\end{figure}

\section{Observation data}
\label{sec:data}

We use the same \xmm dataset (Obs.ID 555200401) analysed in \citet{Furst2011AA}.
The observation was carried out on 2009 July 12, with an exposure of 46 ks, using EPIC-pn in 
the timing mode. The timing mode
allows observations of count rates up to 400 cts~s$^{-1}$ without detectable pile-up.
The corresponding orbital phases are $\sim0.98$, according to the ephemeris of \citet{Doroshenko2010AA}
($\sim0.92$ for the ephemeris of \citet{Koh1997ApJ}).

The data were reprocessed with the Science Analysis Software (SAS 17.0.0), 
following the standard procedures. 
Compared with the data reduction in \citet{Furst2011AA}, the biggest difference is the application 
of a rate-dependent correction of energy scale in the Fe \Ka band for PN exposures 
in timing mode\footnote{http://xmm2.esac.esa.int/docs/documents/CAL-SRN-0312-1-4.pdf}. 
Before filtering and extracting the high level products with \textit{evselevt}, 
the barycentric correction is applied via ftool \textit{barycen}, and the binary effect
is corrected with the program {\it binaryCor} in Remeis 
ISISscripts\footnote{http://www.sternwarte.uni-erlangen.de/isis}.
The source data were extracted from Cols.~29--47 and the background data from 
Cols.~3--5.

\section{analysis and results}
\label{sec:result}

The extracted light curves of the hard continuum (7.8--12 keV) and the Fe \Ka line ($6.25-6.55$~keV) 
are presented in Figure~\ref{fig:Fe_continuum}. 
The energy range of $7.8-12$~keV is chosen to exclude the contributions from Fe \Kb and Ni \Ka lines.
A baseline continuum for the Fe \Ka line is estimated from a linear interpolation between the 
rates within $5.7-6.1$~keV and $6.7-6.9$~keV, and is shown as the navy line in 
Figure~\ref{fig:Fe_continuum}.
The hard continuum shows intense flares at the beginning 
($0-15$~ks) and the ending ($30-46$~ks) of the observation, and 
stays relatively quiescent in the middle ($15-30$~ks).
The Fe \Ka fluxes show
good correlations with the hard continuum fluxes in the quiescent period, but not in the flare periods.
The averaged rate of the interpolated baseline continuum of the Fe \Ka line is about 30\% 
of that of the Fe \Ka line in the middle quiescent period, and is comparable to that of the 
Fe \Ka line during the flare periods.
It illustrates the significant contribution of the underlying continuum to the observed Fe \Ka line.

As the Fe \Ka photons are emitted from the reprocession of the intrinsic hard X-ray continuum
of the neutron star by the surrounding material, a time delay of the Fe \Ka line 
behind the hard continuum, 
corresponding to the average distance of the surrounding material to the neutron star, is expected.
We apply the standard cross-correlation function (CCF) method \citep[e.g.][]{Peterson1993PASP} 
to measure the time delay. The CCF is defined as:

\begin{equation}
\begin{split}
	\label{eq:ccf}
   {\rm  CCF(\tau)=\sum_t L_{c}(t-\tau)~L_{Fe}(t)}
\end{split}
\end{equation}
where $\rm L_{c}$ and  $\rm L_{Fe}$ are the light curves of the hard continuum and 
the Fe \Ka line, respectively. Both $\rm L_{c}$ and  $\rm L_{Fe}$ are normalized by the root 
mean square of the light curves to ensure that the resulting CCF values are between -1 and 1. 

\subsection{CCFs for the whole observation}

We calculated three CCFs with respect to the hard continuum
($7.8-12$~keV): for the interpolated continuum 
in the Fe \Ka band, the continuum-unsubtracted Fe \Ka line, and 
the continuum-subtracted Fe \Ka line, respectively.
The light curves (with a binsize of 5~s) over the whole observation are summed.
The resulting CCFs are presented in Figure~\ref{fig:10sbin3line}.
The CCF between the interpolated continuum and the hard continuum is peaked around 
0 and shows a symmetric profile, indicating that they have almost no time delay. 
The CCF between the continuum-unsubtracted Fe \Ka line and the hard continuum is also
peaked around 0, showing no significant time delay, as reported by \citet{Furst2011AA}. 
Note that its profile is a little skewed toward the positive side, which is an indication
of a positive time delay \citep{Peterson1998PASP}.
The CCF between the continuum-subtracted Fe \Ka line and the hard continuum, however, is peaked 
around 40~s, and is also skewed toward the positive side. 
Therefore, the CCF between the continuum-unsubtracted Fe \Ka line and the hard continuum
can be regarded as a combination of the CCFs of the interpolated continuum and the net Fe \Ka line
with respect to the hard continuum. 
The interpolated underlying continuum dominates the location of the peak of the continuum-unsubtracted 
CCF (0~s), while the Fe \Ka line produces the skewness of the continuum-unsubtracted CCF.
The time delay of the Fe \Ka line with respect to the hard continuum
becomes obvious only when the underlying continuum is subtracted.

We estimated the measurement errors of the time delay by re-sampling the light curves of 
the Fe \Ka line and the hard continuum with a Possion distribution for 5000 times, 
and calculating the distribution of the peaks of sampled CCFs. 
We found a time delay of $40\pm20$~s.

\begin{figure} 
  \includegraphics[width=0.7\columnwidth,angle=270]{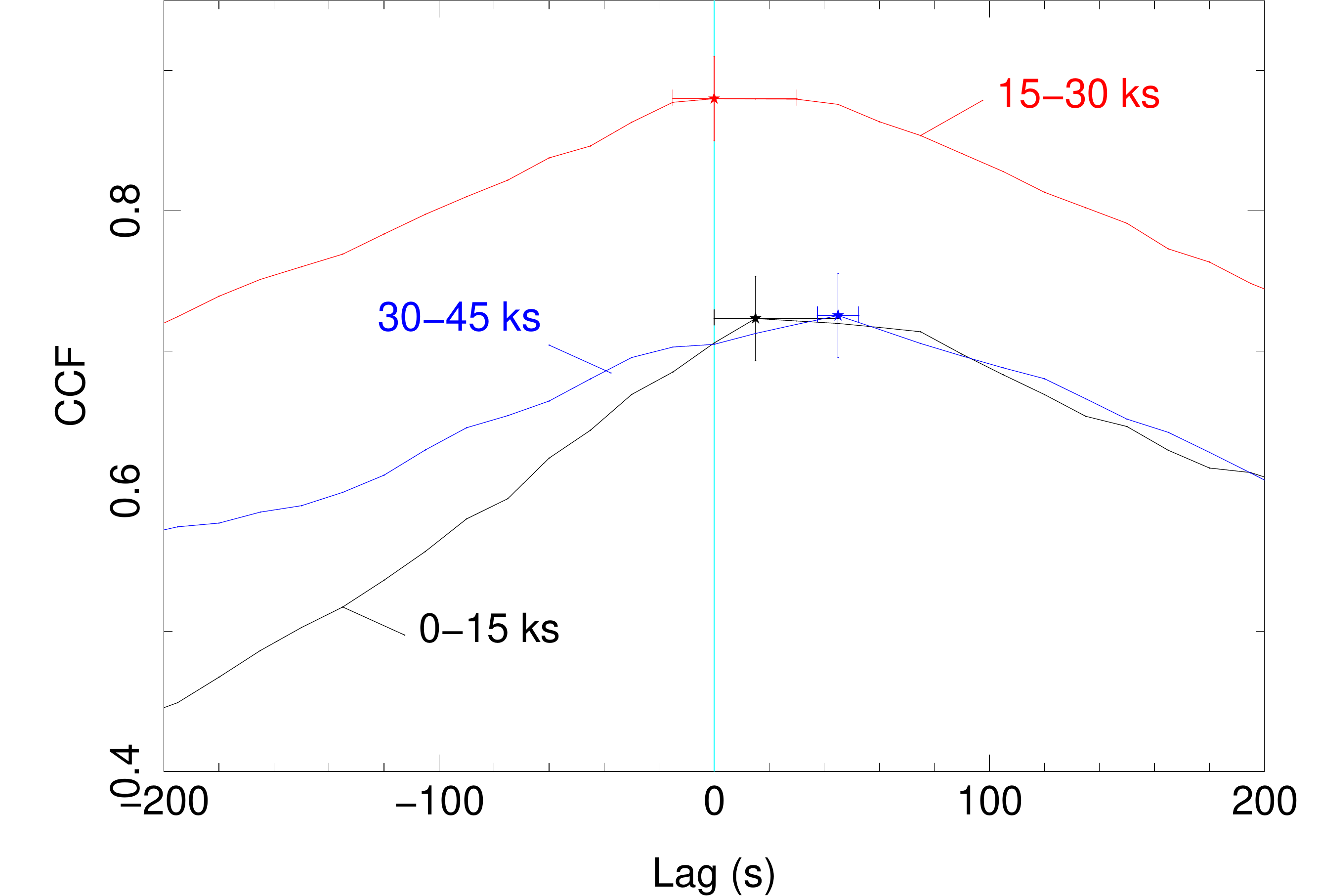} 
  \caption{ CCFs for three 15 ks periods. They indicate that the Fe \KA-emitting region is varying.
}
  \label{fig:15ks_15s}
\end{figure}

\subsection{CCFs for 15~ks periods}
\label{sec:15ks duration}

\begin{figure*}
\centering

\begin{minipage}[]{0.3\textwidth}
\centering
\includegraphics[width=5cm,angle=270]{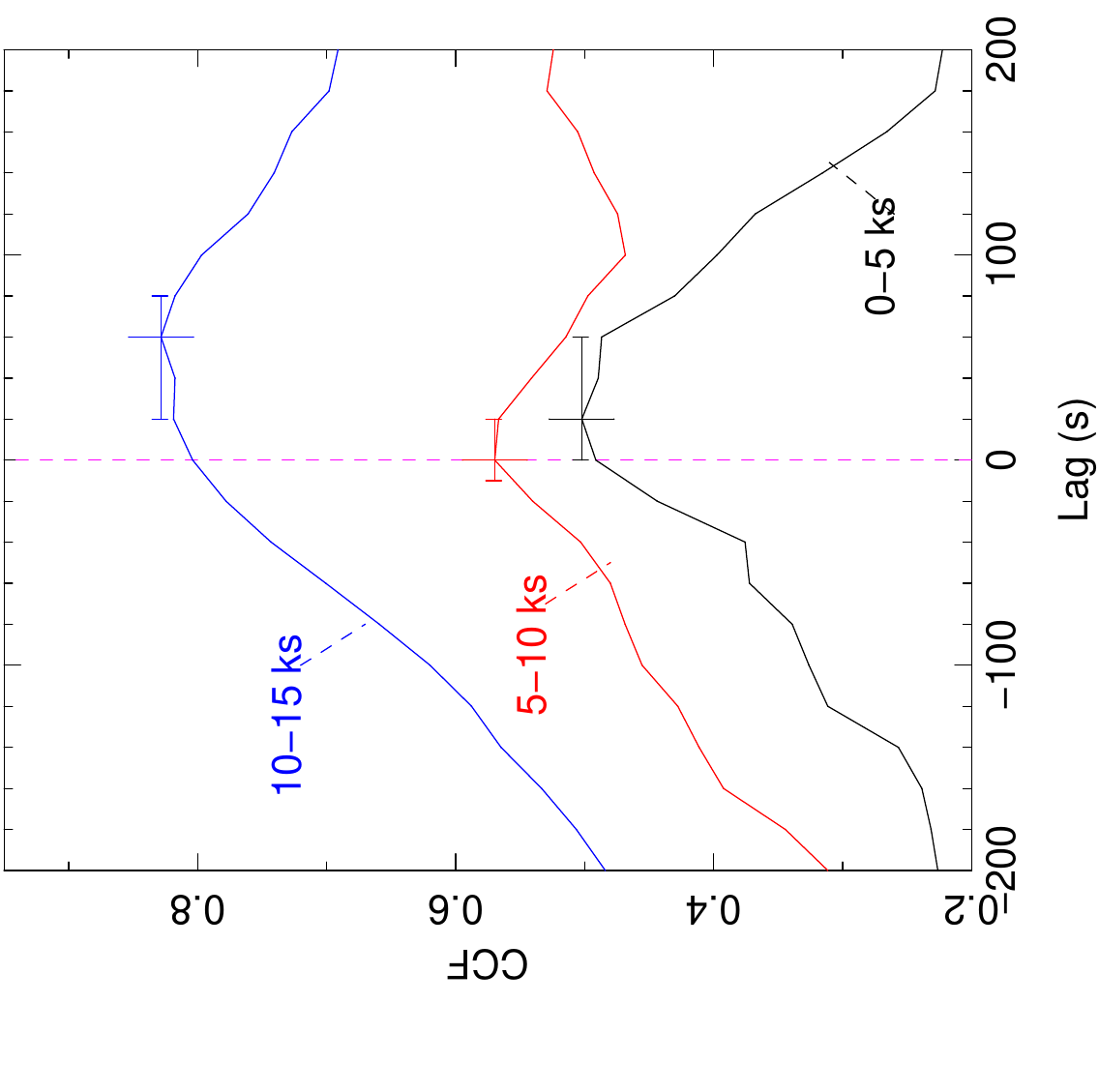}
\end{minipage}%
\begin{minipage}[]{0.3\textwidth}
\centering
\includegraphics[width=5cm,angle=270]{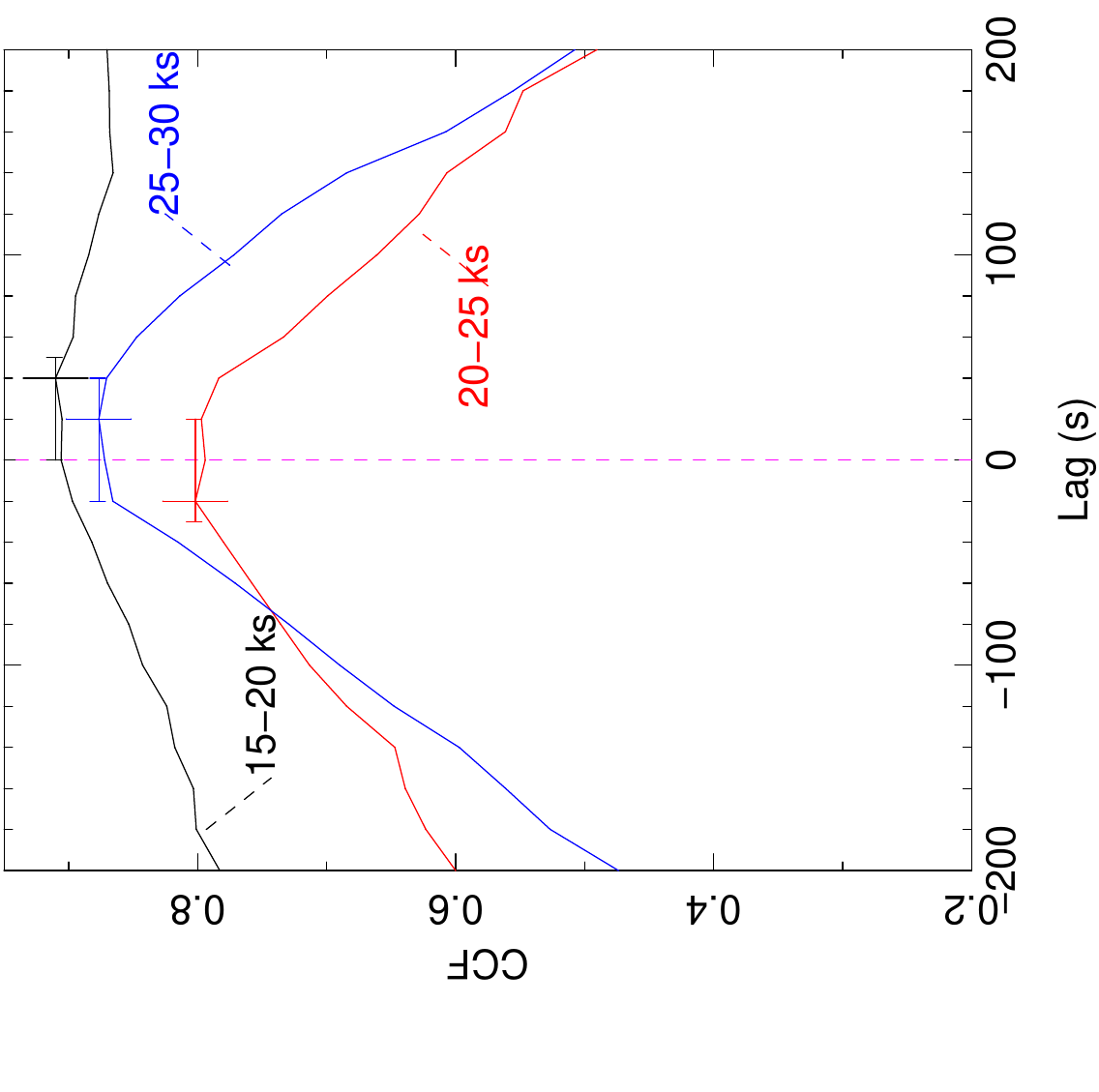}
\end{minipage}
\begin{minipage}[]{0.3\textwidth}
\centering
\includegraphics[width=5cm,angle=270]{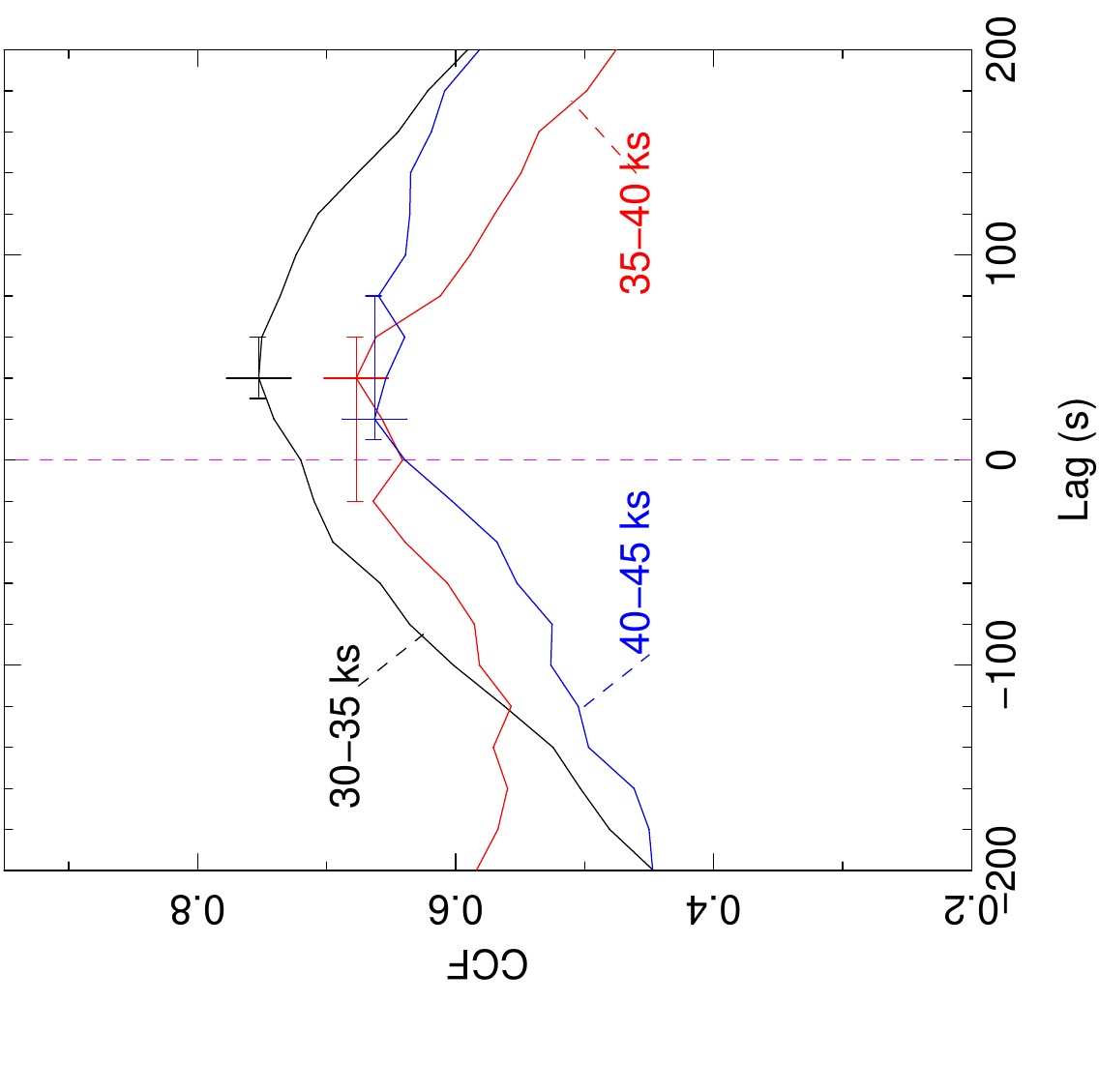}
 \label{fig:30000-45000}
\end{minipage}
\caption{CCFs for 5 ks periods in 20~s resolution. The vertical lines represent the peaks. Left: 0-5, 5--10, 10--15 ks; Middle: 15--20, 20--25, 25--30 ks; Right: 30--35, 35--0, 40--45 ks.}
\label{fig:3panel}
\end{figure*}

As discussed above, the hard continuum of GX 301-2 shows intense flares for the first 15~ks and the last 
15~ks, and keeps quiescent in the middle, and the Fe \Ka fluxes are well correlated with the 
continuum fluxes in the middle quiescent period, but not in the flare periods. To study the 
dependence of the CCF on the flare activity, we divided the observation into three 15~ks periods 
and calculated the CCF between the continuum-subtracted Fe \Ka line and the hard continuum 
($7.8-12$~keV), respectively.
In order to obtain a stable CCF profile, the light curves are binned in 15~s.
The resulting CCFs are plotted in Figure~\ref{fig:15ks_15s}.

The CCF of the middle quiescent period shows the highest correlation value, consistent with 
the apparent good correlation between the Fe \Ka fluxes and the hard continuum fluxes
 shown in Figure~\ref{fig:Fe_continuum}.
The peak of the CCF of the middle 15ks period is located around $0^{+30}_{-15}$~s. The CCFs of the first 15 ks and the last 
15 ks show a relatively low correlation value, and have positive peaks 
at $15^{+30}_{-15}$~s and $45\pm7.5$~s, respectively. 
Because the peaks of sampled CCFs of the last 15~ks are all located in one bin, half of 
the binsize 7.5~s is taken as the errors.
The CCF profile of the first 15 ks shows a clear skewness towards the positive side.
These results indicate that the Fe \KA-emitting region is varying on a time-scale of 15~ks.

\subsection{CCFs for 5 ks periods}
\label{sec:5ks duration}

To further study the variation of the Fe \KA-emitting region, which implying an unsteady 
accretion flow around the neutron star,  
we further derived the three 15 ks periods into smaller time spans of 5 ks.
The light curves are binned in 20~s.
The resulting CCFs are plotted in Figure~\ref{fig:3panel}.
Similar to the cases of 15~ks periods, most of the CCFs have a positive time
delay in the range of $20-60$~s.
During the first 15~ks (the left panel), the intense flare within 10--15 ks stands out for its 
higher correlation value and apparent time lag of $60^{+20}_{-40}$\,s.
During the second 15~ks (the middle panel), the quiescent period of $15-20$~ks shows the 
highest correlation value,
and its profile is more spreading compared to other periods.
During the last 15~ks (the right panel), the intense flare within 30-35 ks is also remarkable for its
higher correlation value and prominent time lag of $40^{+20}_{-10}$\,s.
In general, the intense flare periods ($10-15$~ks and $30-35$~ks) show the largest time delays,
while the quiescent periods show smaller time delays.

\section{Discussion and conclusion}
\label{sec:discussion}

We studied the time delay between the Fe \Ka line and the hard continuum of GX 301-2
using the CCF method with \xmm data. We found that the subtraction of the underlying continuum in 
the Fe \Ka band is crucial to reveal the true time delay between the Fe \Ka line and the 
hard continuum. The measured time delay over the whole observation period is 
$40\pm20$~s.
The measured time delay of the middle quiescent period ($15-30$~ks) is $0^{+30}_{-15}$~s,
a little smaller than those of the flare periods ($0-15$~ks and $30-45$~ks). 
The best measurements of positive time delays are from the intense flare periods 
of $10-15$~ks and $30-35$~ks, which are $60^{+20}_{-40}$~s and $40^{+20}_{-10}$~s, respectively.
These results show that the typical Fe \KA-emitting region of GX 301-2 is around 40~lt-s, and 
it can vary over the time-scale of $5-15$~ks.

Our results are quite different from those reported previously for GX 301-2.
\citet{Endo2002ApJ} inferred a smallest Fe \KA-emitting region $\sim0.3$~lt-s.
It is most likely because they over-estimated the Fe \Ka line width due to the limited 
spectral resolution of \ascato. With an higher spectral resolution, \cha HETG observation of 
GX 301-2 provided a Fe \Ka line width $\sim 4$~eV 
\citep[e.g.][]{Liu2018MNRAS,Tzanavaris2018ApJ}, indicating a much larger scale.
\citet{Furst2011AA} found no significant time delay beyond 2~s, which, as shown in the analysis of 
\S~\ref{sec:result}, is because that the underlying continuum within the Fe \Ka band was not subtracted.
They also identified a period of low state (lasting for $\sim4.5$~ks) and 
inferred a second absorber on 2000~lt-s scale.
The time delay during this low state ($17-21$~ks) is $\sim 0$~s, inconsistent with 
an emitting region of a large scale of 2000~lt-s.
A possible scenario is that a thick gas structure temporarily 
obscured the intrinsic hard emission of the neutron star, and the observed hard continuum were mainly
from scattered emission. To heavily obscure the photons of $7.8-12$~keV, 
a column density larger than $2\times10^{24}$~cm$^{-2}$ is required. In principle, 
the Fe \KA-emitting gas itself could play the absorption role.

On the other hand, \citet{Suchy2012ApJ} inferred a scale larger than 700~lt-s based on 
the flat pulse profile of the Fe \Ka line. We note that as long as the fluorescence material 
is quasi-symmetric with respect to the rotational axis of the neutron star, the illuminating 
radiation on the fluorescence material will be similar for any spin phases, and the resulting 
Fe \Ka line will show no apparent pulsation. 
The fluorescence material is not necessarily to be as far as 
the light travel distance of spin period (685~lt-s) to destroy the coherence.

The circumstellar material will be gravitationally bounded by the neutron star if it is within 
the accretion radius:
\begin{equation}
\begin{split}
   \label{eq:quadratic}
\rm
    R_{acc}&=\frac{2GM}{\rm v_{rel}^2}\simeq55\frac{M}{1.4~M_\odot}(\rm \frac{v_{rel}}{\rm 150 ~km~s^{-1}})^{-2}{\rm~\hbox{lt-s}},\\
\end{split}
\end{equation}

where $\rm v_{rel}=\sqrt{v_{orb}^2+v^2_w}$ is the relative velocity of the neutral star 
with respect to the surrounding material, $\rm v_{orb}$ is the orbital velocity, and $\rm v_w$ is the wind
velocity at the location of the neutron star.
$\rm v_{orb}\sim350$~\kms~and $\rm v_w\sim150$~\kms~near the periastron 
adopting the orbit parameters of \citet{Doroshenko2010AA}
and a terminal velocity of $400$~\kms. These values provide an accretion radius $\rm R_{acc}\sim8$~lt-s, 5 
times less than the measured average size of Fe \KA-emitting region.

A very particular aspect of GX 301-2 is that it shows pre-periastron X-ray flares
around orbital phases $\sim0.92$ \citep[e.g.][]{Leahy2002AA,Islam2014MNRAS}. It could be 
explained by an accretion stream originating from the surface of the donor star
\citep{Leahy2008MNRAS}. 
They inferred that the stream has a mass-loss rate 2-3 times larger than that of the stellar wind.
Thus the stream could be the dominate mass-loss mechanism of the massive star of GX 301-2.
Evidences for the existence of the stream are also 
found in the studies of optical lines \citep{Kaper2006AA} and mid-infrared interferometry 
\citep{Waisberg2017ApJ}.
If the accretion stream is indeed the dominant mass-loss mechanism,
the neutron star is passing the stream near the periastron, and the relative velocity between 
the stream and the neutron star in the orbital direction will be small. In this case, 
$\rm R_{acc}\sim55$~lt-s, similar to the measured average size of Fe \KA-emitting region.
Since $\rm R_{acc}$ is a natural physical scale for the distribution of the 
circumstellar material,
the measured Fe \KA-emitting region favours the case of the accretion stream,
rather than the quasi-isotropic stellar wind.

It is very interesting to note that the measured sizes of Fe \KA-emitting region on 
time-scales of 5-15~ks are variable, changing from a larger scale during the intense flares, 
to a smaller scale for the quiescent periods. Such a behavior is consistent with the 
unsteady nature of the accretion flow in wind-fed systems, as illustrated in numerical
simulations \citep[e.g.][ among others]{Taam1988ApJ,Blondin1990ApJ}.
A change of 10~lt-s on time-scales of 10~ks implies a velocity of 300~\kms, which is physically plausible.
The significance of the unsteady behaviour of the Fe \KA-emitting material is limited by the current
observation data, and further observations of GX 301-2 will help to improve the significance and to 
probe the unsteady evolution of the circumstellar material on longer time-scales.

In conclusion, 
these results show that the spatial distribution of the circumstellar material of GX 301-2
can be measured from the time delay between the Fe \Ka line and the hard continuum using
CCF method. This methodology might be applicable to other few SgXBs with bright 
Fe \Ka lines using archival data, and will be explored in a future work. The available candidates
would be limited to few bright sources. Nevertheless,
with future X-ray missions of larger collecting area, such as Athena and eXTP,
much more SgXBs would be available to be studied using the CCF method, as illustrated here. 
Statistical studies of the circumstellar material with other properties in these systems, 
such as wind velocities, are expected. The unsteady behaviour of the accretion flow in such systems 
will be better revealed, and its correlation with other properties, such as the observed fluxes,
will be studied in great details.

\section*{Acknowledgements}
JL acknowledges the support by National Natural Science Foundation of China (NSFC, 11773035). LG acknowledges the supported by the National Program on Key Research and Development Project through grant No. 2016YFA0400804, and by the National NSFC with grant No. Y913041V01, and by the Strategic Priority Research Program of the Chinese Academy of Sciences through grant No. XDB23040100. This work has also made use of the archive data from the XMM-Newton, an ESA science mis- sion funded by ESA Member States and USA (NASA).

\bibliographystyle{mnras}
\bibliography{lag1106} 

\begin{thebibliography}{}
\makeatletter
\relax
\def\mn@urlcharsother{\let\do\@makeother \do\$\do\&\do\#\do\^\do\_\do\%\do\~}
\def\mn@doi{\begingroup\mn@urlcharsother \@ifnextchar [ {\mn@doi@}
  {\mn@doi@[]}}
\def\mn@doi@[#1]#2{\def\@tempa{#1}\ifx\@tempa\@empty \href
  {http://dx.doi.org/#2} {doi:#2}\else \href {http://dx.doi.org/#2} {#1}\fi
  \endgroup}
\def\mn@eprint#1#2{\mn@eprint@#1:#2::\@nil}
\def\mn@eprint@arXiv#1{\href {http://arxiv.org/abs/#1} {{\tt arXiv:#1}}}
\def\mn@eprint@dblp#1{\href {http://dblp.uni-trier.de/rec/bibtex/#1.xml}
  {dblp:#1}}
\def\mn@eprint@#1:#2:#3:#4\@nil{\def\@tempa {#1}\def\@tempb {#2}\def\@tempc
  {#3}\ifx \@tempc \@empty \let \@tempc \@tempb \let \@tempb \@tempa \fi \ifx
  \@tempb \@empty \def\@tempb {arXiv}\fi \@ifundefined
  {mn@eprint@\@tempb}{\@tempb:\@tempc}{\expandafter \expandafter \csname
  mn@eprint@\@tempb\endcsname \expandafter{\@tempc}}}

\bibitem[\protect\citeauthoryear{{Blondin}, {Kallman}, {Fryxell}  \&
  {Taam}}{{Blondin} et~al.}{1990}]{Blondin1990ApJ}
{Blondin} J.~M.,  {Kallman} T.~R.,  {Fryxell} B.~A.,   {Taam} R.~E.,  1990,
  \mn@doi [\apj] {10.1086/168865}, \href
  {https://ui.adsabs.harvard.edu/abs/1990ApJ...356..591B} {356, 591}

\bibitem[\protect\citeauthoryear{{Davidson} \& {Ostriker}}{{Davidson} \&
  {Ostriker}}{1973}]{Davidson1973ApJ}
{Davidson} K.,  {Ostriker} J.~P.,  1973, \mn@doi [\apj] {10.1086/151897}, \href
  {https://ui.adsabs.harvard.edu/abs/1973ApJ...179..585D} {179, 585}

\bibitem[\protect\citeauthoryear{{Doroshenko}, {Santangelo}, {Suleimanov},
  {Kreykenbohm}, {Staubert}, {Ferrigno}  \& {Klochkov}}{{Doroshenko}
  et~al.}{2010}]{Doroshenko2010AA}
{Doroshenko} V.,  {Santangelo} A.,  {Suleimanov} V.,  {Kreykenbohm} I.,
  {Staubert} R.,  {Ferrigno} C.,   {Klochkov} D.,  2010, \mn@doi [\aap]
  {10.1051/0004-6361/200912951}, \href
  {https://ui.adsabs.harvard.edu/abs/2010A&A...515A..10D} {515, A10}

\bibitem[\protect\citeauthoryear{{Endo}, {Ishida}, {Masai}, {Kunieda}, {Inoue}
  \& {Nagase}}{{Endo} et~al.}{2002}]{Endo2002ApJ}
{Endo} T.,  {Ishida} M.,  {Masai} K.,  {Kunieda} H.,  {Inoue} H.,   {Nagase}
  F.,  2002, \mn@doi [\apj] {10.1086/341060}, \href
  {https://ui.adsabs.harvard.edu/abs/2002ApJ...574..879E} {574, 879}

\bibitem[\protect\citeauthoryear{{F{\"u}rst} et~al.,}{{F{\"u}rst}
  et~al.}{2011}]{Furst2011AA}
{F{\"u}rst} F.,  et~al., 2011, \mn@doi [\aap] {10.1051/0004-6361/201117665},
  \href {https://ui.adsabs.harvard.edu/abs/2011A&A...535A...9F} {535, A9}

\bibitem[\protect\citeauthoryear{{Gim{\'e}nez-Garc{\'\i}a}, {Torrej{\'o}n},
  {Eikmann}, {Mart{\'\i}nez-N{\'u}{\~n}ez}, {Oskinova}, {Rodes-Roca}  \&
  {Bernab{\'e}u}}{{Gim{\'e}nez-Garc{\'\i}a} et~al.}{2015}]{Gimenez2015AA}
{Gim{\'e}nez-Garc{\'\i}a} A.,  {Torrej{\'o}n} J.~M.,  {Eikmann} W.,
  {Mart{\'\i}nez-N{\'u}{\~n}ez} S.,  {Oskinova} L.~M.,  {Rodes-Roca} J.~J.,
  {Bernab{\'e}u} G.,  2015, \mn@doi [\aap] {10.1051/0004-6361/201425004}, \href
  {https://ui.adsabs.harvard.edu/abs/2015A&A...576A.108G} {576, A108}

\bibitem[\protect\citeauthoryear{{Haberl}, {White}  \& {Kallman}}{{Haberl}
  et~al.}{1989}]{Haberl1989ApJ}
{Haberl} F.,  {White} N.~E.,   {Kallman} T.~R.,  1989, \mn@doi [\apj]
  {10.1086/167714}, \href
  {https://ui.adsabs.harvard.edu/abs/1989ApJ...343..409H} {343, 409}

\bibitem[\protect\citeauthoryear{{Islam} \& {Paul}}{{Islam} \&
  {Paul}}{2014}]{Islam2014MNRAS}
{Islam} N.,  {Paul} B.,  2014, \mn@doi [\mnras] {10.1093/mnras/stu756}, \href
  {https://ui.adsabs.harvard.edu/abs/2014MNRAS.441.2539I} {441, 2539}

\bibitem[\protect\citeauthoryear{{Kaper}, {van der Meer}  \& {Najarro}}{{Kaper}
  et~al.}{2006}]{Kaper2006AA}
{Kaper} L.,  {van der Meer} A.,   {Najarro} F.,  2006, \mn@doi [\aap]
  {10.1051/0004-6361:20065393}, \href
  {https://ui.adsabs.harvard.edu/abs/2006A&A...457..595K} {457, 595}

\bibitem[\protect\citeauthoryear{{Koh} et~al.,}{{Koh}
  et~al.}{1997}]{Koh1997ApJ}
{Koh} D.~T.,  et~al., 1997, \mn@doi [\apj] {10.1086/303929}, \href
  {https://ui.adsabs.harvard.edu/abs/1997ApJ...479..933K} {479, 933}

\bibitem[\protect\citeauthoryear{{Leahy}}{{Leahy}}{2002}]{Leahy2002AA}
{Leahy} D.~A.,  2002, \mn@doi [\aap] {10.1051/0004-6361:20020781}, \href
  {https://ui.adsabs.harvard.edu/abs/2002A&A...391..219L} {391, 219}

\bibitem[\protect\citeauthoryear{{Leahy} \& {Kostka}}{{Leahy} \&
  {Kostka}}{2008}]{Leahy2008MNRAS}
{Leahy} D.~A.,  {Kostka} M.,  2008, \mn@doi [\mnras]
  {10.1111/j.1365-2966.2007.12754.x}, \href
  {https://ui.adsabs.harvard.edu/abs/2008MNRAS.384..747L} {384, 747}

\bibitem[\protect\citeauthoryear{{Liu}, {Soria}, {Qiao}  \& {Liu}}{{Liu}
  et~al.}{2018}]{Liu2018MNRAS}
{Liu} J.,  {Soria} R.,  {Qiao} E.,   {Liu} J.,  2018, \mn@doi [\mnras]
  {10.1093/mnras/sty2180}, \href
  {https://ui.adsabs.harvard.edu/abs/2018MNRAS.480.4746L} {480, 4746}

\bibitem[\protect\citeauthoryear{{Mart{\'\i}nez-N{\'u}{\~n}ez}
  et~al.,}{{Mart{\'\i}nez-N{\'u}{\~n}ez} et~al.}{2017}]{Martinez-Nunez2017SSRv}
{Mart{\'\i}nez-N{\'u}{\~n}ez} S.,  et~al., 2017, \mn@doi [\ssr]
  {10.1007/s11214-017-0340-1}, \href
  {https://ui.adsabs.harvard.edu/abs/2017SSRv..212...59M} {212, 59}

\bibitem[\protect\citeauthoryear{{Nagase}}{{Nagase}}{1989}]{Nagase1989PASJ}
{Nagase} F.,  1989, \pasj, \href
  {https://ui.adsabs.harvard.edu/abs/1989PASJ...41....1N} {41, 1}

\bibitem[\protect\citeauthoryear{{Peterson}}{{Peterson}}{1993}]{Peterson1993PASP}
{Peterson} B.~M.,  1993, \mn@doi [\pasp] {10.1086/133140}, \href
  {https://ui.adsabs.harvard.edu/abs/1993PASP..105..247P} {105, 247}

\bibitem[\protect\citeauthoryear{{Peterson}, {Wanders}, {Horne}, {Collier},
  {Alexander}, {Kaspi}  \& {Maoz}}{{Peterson} et~al.}{1998}]{Peterson1998PASP}
{Peterson} B.~M.,  {Wanders} I.,  {Horne} K.,  {Collier} S.,  {Alexander} T.,
  {Kaspi} S.,   {Maoz} D.,  1998, \mn@doi [\pasp] {10.1086/316177}, \href
  {https://ui.adsabs.harvard.edu/abs/1998PASP..110..660P} {110, 660}

\bibitem[\protect\citeauthoryear{{Suchy}, {F{\"u}rst}, {Pottschmidt},
  {Caballero}, {Kreykenbohm}, {Wilms}, {Markowitz}  \& {Rothschild}}{{Suchy}
  et~al.}{2012}]{Suchy2012ApJ}
{Suchy} S.,  {F{\"u}rst} F.,  {Pottschmidt} K.,  {Caballero} I.,  {Kreykenbohm}
  I.,  {Wilms} J.,  {Markowitz} A.,   {Rothschild} R.~E.,  2012, \mn@doi [\apj]
  {10.1088/0004-637X/745/2/124}, \href
  {https://ui.adsabs.harvard.edu/abs/2012ApJ...745..124S} {745, 124}

\bibitem[\protect\citeauthoryear{{Taam} \& {Fryxell}}{{Taam} \&
  {Fryxell}}{1988}]{Taam1988ApJ}
{Taam} R.~E.,  {Fryxell} B.~A.,  1988, \mn@doi [\apjl] {10.1086/185143}, \href
  {https://ui.adsabs.harvard.edu/abs/1988ApJ...327L..73T} {327, L73}

\bibitem[\protect\citeauthoryear{{Torrej{\'o}n}, {Schulz}, {Nowak}  \&
  {Kallman}}{{Torrej{\'o}n} et~al.}{2010}]{Torrejon2010ApJ}
{Torrej{\'o}n} J.~M.,  {Schulz} N.~S.,  {Nowak} M.~A.,   {Kallman} T.~R.,
  2010, \mn@doi [\apj] {10.1088/0004-637X/715/2/947}, \href
  {https://ui.adsabs.harvard.edu/abs/2010ApJ...715..947T} {715, 947}

\bibitem[\protect\citeauthoryear{{Tzanavaris} \& {Yaqoob}}{{Tzanavaris} \&
  {Yaqoob}}{2018}]{Tzanavaris2018ApJ}
{Tzanavaris} P.,  {Yaqoob} T.,  2018, \mn@doi [\apj]
  {10.3847/1538-4357/aaaab6}, \href
  {https://ui.adsabs.harvard.edu/abs/2018ApJ...855...25T} {855, 25}

\bibitem[\protect\citeauthoryear{{Waisberg} et~al.,}{{Waisberg}
  et~al.}{2017}]{Waisberg2017ApJ}
{Waisberg} I.,  et~al., 2017, \mn@doi [\apj] {10.3847/1538-4357/aa79f1}, \href
  {https://ui.adsabs.harvard.edu/abs/2017ApJ...844...72W} {844, 72}

\bibitem[\protect\citeauthoryear{{Watanabe} et~al.,}{{Watanabe}
  et~al.}{2003}]{Watanabe2003ApJ}
{Watanabe} S.,  et~al., 2003, \mn@doi [\apjl] {10.1086/379735}, \href
  {https://ui.adsabs.harvard.edu/abs/2003ApJ...597L..37W} {597, L37}

\makeatother
\end{thebibliography}

\appendix

\bsp	
\label{lastpage}
\end{document}